**Manuscript title:** **Deep Learning-based Multiparametric Quantitative Ultra-sound for Hepatic Steatosis Assessment**

**Authors:** Allison Lee[1], Ahmed El Kaffas[2], Aya Kamaya[3], Jeremy J. Dahl[3], Jihye Baek[3]

[1]Department of Electrical Engineering, Stanford University, Palo Alto, CA, USA

[2]Department of Radiology, University of California San Diego, La Jolla, CA, USA

[3]Department of Radiology, Stanford University, Palo Alto, CA, 94304, USA

**Corresponding author:** Jihye Baek

Stanford University

3155 Porter Dr., Palo Alto, CA 94304

Phone: 585-568-9804

Email: baekjh@stanford.edu

**Abstract**

Quantitative ultrasound (QUS) has shown promise for liver fat quantification in metabolic dysfunction-associated fatty liver disease (MAFLD). Incorporating multiple parameters has demonstrated the potential to outperform single QUS parameter measures. Here, we propose a deep learning-based multiparametric QUS framework that learns effective integration of multiple features from QUS measures to estimate pixel-wise hepatic fat fraction, enabling the detecting of heterogeneous fat accumulation within the liver.

We developed a U-Net based network architecture to predict local fat fraction percentages, which takes 2D QUS maps as input and is trained with magnetic resonance imaging-proton density fat fraction (MRI-PDFF) as the ground truth. We extracted 2D QUS maps (H-scan frequency, attenuation, Nakagami, and RF/envelope/log-compressed data) and tested various input configurations consisting of different combinations of the QUS parameters. We enrolled 30 patients with healthy livers or suspected/confirmed MASLD, and acquired ultrasound RF data, and validated our approach using the patient data.

Among all tested input configurations, the model combining H-scan, attenuation, and Nakagami parameters achieved the best performance, with a Spearman correlation coefficient ($R_s$) of 0.92 and an area under the curve (AUC) of 0.98. The predicted fat fraction maps exhibited color changes concordant with the gradual increasing in MRI-PDFF, from normal liver to 45% fat accumulation. These findings demonstrate that deep learning can integrate multiparametric QUS information in a fully data-driven manner, providing a localized fat fraction map rather than a single whole-liver value. We anticipate that our deep learning framework can offer potential advantages for future assessment of heterogeneous disease development in liver for more complex, progressive liver diseases such as steatohepatitis.

## 1. Introduction

Metabolic dysfunction-associated steatotic liver disease (MASLD) is rising in prevalence worldwide and is now a leading cause of chronic liver disease [1-3]. While affecting up to 30% of adults globally [1] and projected to increase by 56% in the coming decades [2], MASLD is the fastest-growing liver disease [3]. MASLD begins with hepatic steatosis, which can progress to metabolic dysfunction-associated steatohepatitis (MASH), a more severe form defined by the co-existence of steatosis, inflammation, and hepatocellular ballooning, with or without fibrosis [4-7]. MASH is associated with an increased risk of adverse outcomes, including cardiovascular diseases, liver-related complications, and mortality in patients with significant fibrosis [4, 8, 9]. Because hepatic steatosis is the defining feature of MASLD and precedes its progressive stages, early detection and longitudinal monitoring of steatosis are essential for timely intervention, better patient outcomes, and lower healthcare costs [10].

Liver biopsy is currently the reference standard for diagnosing hepatic steatosis, but it is invasive, prone to sampling error, and carries risks of complications such as bleeding and infection [11]. Noninvasive imaging modalities have therefore been investigated as alternatives. Magnetic resonance imaging–derived proton density fat fraction (MRI-PDFF) is widely accepted as an accurate noninvasive reference for hepatic fat quantification [12-14]. However, despite its quantitative measurements, the high cost and limited availability of MRI restrict its use for broad screening and repeated monitoring of liver diseases. In contrast, ultrasound is widely available and relatively inexpensive, and supports real-time imaging, making it well suited for longitudinal monitoring of disease progression as well as detection of disease onset.

Conventional B-mode ultrasound assessment relies on visually interpreting qualitative features, such as increased liver echogenicity and reduced visualization of deeper structures caused by greater acoustic attenuation, to detect increasing fat accumulation in the liver. However, such visual screening is inherently qualitative and operator-dependent, leading to limited reproducibility. This limited reliability has motivated the development of quantitative approaches that extract tissue information from ultrasound signals, such as scattering, attenuation, and speckle statistics, yielding reproducible measurements that enable consistent interpretation across operators and examinations. Various QUS parameters have been investigated for hepatic fat assessment. Measures based on attenuation, the hepatorenal index (HRI), Nakagami, and QUS spectroscopy (spectral intercept (SI), spectral slope (SS), and midband fit (MBF)) have shown promise in detecting hepatic steatosis [15]. Moreover, H-scan analysis [16, 17] as an alternative spectral approach was applied for hepatic fat assessment, yielding improved correlation with MRI-PDFF compared to spectroscopy for the same patient dataset [15, 18]. To further improve diagnostic performance over single-parameter measures, prior studies conducted multiparametric QUS approaches, such as principal component analysis (PCA) [18, 19] and nomogram [20], demonstrating the advantages of combining complementary measurements. However, these multiparametric approaches typically yielded a single score per whole liver. Although an extension toward localized imaging based on the whole-liver analysis was performed, it provided limited spatial resolution [18], which may not adequately capture the spatial heterogeneity of fat accumulation.

Deep learning offers a framework for learning relationships between ultrasound inputs and reference steatosis assessments in a data-driven manner. For example, a convolutional neural network (CNN) was used for steatosis staging [21] with different ultrasound image representations, including radiofrequency (RF), envelope, and log-compressed data, as well as spectral and phase representations, and achieved high areas under the curve (AUCs) above 0.98 for steatosis stages from 0 to 3 defined by the corresponding MRI-PDFF cutoffs. However, stage classification provides only categorical information. Predicting the fat fraction itself may be more clinically useful because it is a continuous measure that can quantify liver fat on a finer scale and capture subtle changes over time. Furthermore, because fat accumulation in MASLD can be heterogeneous, a spatially resolved fat fraction map could provide more information than a single whole-liver value.

In this study, we propose a deep learning-based multiparametric QUS framework (DeepQUS) for hepatic steatosis assessment that predicts the fat fraction at pixel-wise resolution as well as a whole-liver score. Specifically, a U-Net-based network receives individual ultrasound representations or selected combinations of spatial QUS maps as input and produces a spatially resolved fat fraction prediction, using MRI-PDFF as the ground truth. We compared input configurations to determine how channel composition affected predictive performance, with particular attention to the integration of H-scan, attenuation, and Nakagami maps. Fat fraction estimates were evaluated against MRI-PDFF, and model-generated maps were used to visualize the spatial distribution of fat accumulation.

## 2. Methods

### A. Study Design

We enrolled the previously acquired human liver ultrasound dataset, used in [15, 18, 21]. Of the 211 patients screened at Stanford University Medical Center, 30 met the eligibility and data-availability requirements and were included in this study. The cohort included individuals with healthy livers or suspected or confirmed MASLD, with at least one associated risk factor of obesity, diabetes, or hypertension. Other liver or kidney diseases, including hepatic fibrosis and renal disease, were excluded. This study was approved by the Stanford University School of Medicine institutional review board and conducted in accordance with the Health Insurance Portability and Accountability Act.

Ultrasound data were acquired using a Philips EPIQ 7 system (Philips Healthcare, Bothell, WA, USA) equipped with a C5-1 convex-array transducer. Only liver acquisitions were included in our analysis. RF data were stored for offline processing and generation of the ultrasound representations (envelope and log-compressed data) and QUS maps. The preprocessing pipeline used an RF sampling frequency of 20 MHz and a transmit center frequency of 3.2 MHz.

MRI-PDFF served as the reference measurement for model training and evaluation. MRI examinations were performed within 14 days of ultrasound acquisition using a 3.0 T Discovery MR750 scanner (GE Medical Systems, Waukesha, WI, USA) and a six-echo, three-dimensional spoiled gradient-recalled echo sequence using IDEAL-IQ. A radiologist placed one region of interest in each of segments 5–8 of the right hepatic lobe. The four measurements were averaged to obtain the reference MRI-PDFF for each patient. Reference values ranged from 1.25% to 42.8%, with a mean ± standard deviation of 14.1% ± 11.3%.

**B. Ultrasound Representations and QUS Map Generation**

The proposed framework integrates spatial QUS maps to estimate hepatic fat fraction; H-scan, attenuation, and Nakagami maps were generated. RF, envelope, and log-compressed representations were also prepared from the same liver acquisitions for the input-channel comparison experiments.

H-scan maps were generated using the matched-filtering approach [17]. In this approach, attenuation compensation is applied to RF data before processing with a bank of 256 Gaussian filters spanning peak frequencies of 1.5–4.5 MHz. At each spatial location, the filter producing the maximum response determines the local H-scan frequency, providing a frequency-based representation of tissue scattering characteristics. For the present analysis, the H-scan maps were spatially smoothed using a two-dimensional median filter using the kernel size of 7.5 $\lambda \times 7.5\ \lambda$, where $\lambda$ is the ultrasound wavelength.

Attenuation estimation was conducted by the frequency-based approach [16]. Unlike the attenuation-compensated processing used for H-scan, attenuation estimation uses uncompensated RF data to characterize the depth-dependent frequency shift in spectral content. These changes provide an estimate of ultrasound attenuation within the liver.

Nakagami parameter *m* was calculated [22]. The Nakagami distribution characterizes the statistical properties of ultrasound echo-envelope amplitudes, with its shape parameter providing a quantitative descriptor of speckle statistics. In this study, the maps were low pass filtered using a normalized Gaussian kernel using a window size of $10 \times 10\ \lambda$.

In addition to the QUS maps, RF, envelope, and log-compressed B-mode representations were prepared for the input-comparison experiments.

### C. Preprocessing and Training Sample Generation

The ultrasound representations and QUS maps were processed using corresponding liver masks, which were manually contoured, to guide training sample extraction. Binary masks identified candidate liver regions. Samples were divided into five subsets for cross validation.

Following the parameter extraction described in Section II-A, the representations, QUS maps, and masks were downsampled along the axial direction by retaining every fifteenth row, beginning with the first row. Lateral sampling was unchanged. The same sampling interval was applied across all corresponding arrays to preserve spatial correspondence among input channels.

This reduced the disparity between the axial and lateral array dimensions, producing approximately square arrays before patch extraction.

Training samples were generated through random patch extraction at varying locations and sizes within the manual liver contour. For each scan, the size of each patch was randomly selected from the integers 50–250, and candidate starting coordinates were randomly selected within the bounding region of the downsampled liver mask. Candidate regions were checked against the mask before intensity-based quality assessment. Quality criteria were evaluated using the log-envelope representation before flooring at −55 dB. A patch was accepted when fewer than 7% of its pixels had amplitudes below −55 dB and fewer than 0.3% had amplitudes above −20 dB. Corresponding patches were extracted from all available channels using identical spatial coordinates. The procedure targeted five accepted patches per scan, with a maximum of one million candidate iterations.

Each extracted patch was interpolated onto a common 250 × 250-pixel grid using bivariate spline interpolation. Source and destination coordinates were defined over the interval from zero to one in both spatial dimensions. This resizing standardized the network input dimensions while retaining correspondence among channels.

Channel-specific z-score normalization was subsequently applied to account for differences in the numerical scales of the ultrasound representations and QUS maps. For each channel, the mean and population standard deviation were calculated by pooling all pixels across the resized patches included in the normalization dataset.

The normalized representations were assembled into input tensors of dimensions ($250 \times 250 \times C$), where $C$ denotes the number of channels in the evaluated configuration. Input tensors were stored as 32-bit floating-point arrays, and each patch retained its corresponding MRI-PDFF reference value for target construction.

**D. Neural Network Architecture**

We introduce a neural network based on U-Net [23], implemented in TensorFlow [24], which takes 2D QUS maps. The network accepted inputs of dimensions $250 \times 250 \times C$, where $C$ denotes the number of input channels. The encoder–decoder architecture incorporated skip connections to combine features across spatial resolutions.

The encoder comprised four convolutional levels, each applying a 2D convolution, batch normalization, and ReLU activation, with the number of filters doubling at each level and max pooling for downsampling between levels. The decoder symmetrically restored resolution using bilinear upsampling and skip connections from the encoder. The output was then resized to the input dimensions and passed through a $1 \times 1$ convolution with softmax activation to produce a two-channel output.

**E. Ground Truth and Loss Function**

MRI-PDFF was used as the ground truth for training. Each patient has a single MRI-PDFF value averaged across the measures in 5-8 segments in the liver, rather than a spatially registered pixelwise measurement. Thus, each ultrasound patch was assigned the reference MRI-PDFF value of its corresponding patient. This value was replicated across a $250 \times 250$ tensor, producing a spatially constant target for each patch. Target values were normalized by 45%, with values exceeding this limit capped at one.

We implemented a hybrid loss function combining cross entropy and root mean square error (RMSE) with equal weights. The cross entropy compared the prediction and ground truth in a pixel-wise manner, whereas the RMSE compared the single reference MRI-PDFF value with the globally averaged prediction within each patch. For a two-channel output, the loss used the second channel and retained its continuous values without thresholding.

**F. Inference and Fat Fraction Map Generation**

During inference, the whole-liver ultrasound scans were prepared using the channels corresponding to each training set input configuration. Each channel was resampled onto a 250×250-pixel grid using bivariate spline interpolation and standardized using the saved channel-specific means and standard deviations for Z-score normalization. The normalized channels were assembled into a multichannel tensor and supplied to the trained U-Net. Thus, the network was trained using extracted patches and applied to resized full-scan inputs during inference.

The network generated a normalized fat fraction prediction map with continuous values ranging from 0 to 1. These normalized predictions were converted to fat fraction percentages by multiplying by 45, since the training data were normalized using the fat fraction of 45%. The liver contour defined the region used for visualization and quantitative assessment for our analysis.

For quantitative evaluation, predicted fat fraction values were averaged within the liver contour to obtain one estimate per sample. Sample-level estimates were then averaged for each patient and compared with the corresponding reference MRI-PDFF. For visualization, predicted fat fraction values were displayed as color-coded maps and overlaid on the corresponding B-mode images

## 3. Results

**Table 1** summarizes the performance of individual and combined QUS input channels for predicting liver fat fraction using our deep learning framework, compared with the reference MRI-PDFF. We investigated our network performance for the input configurations listed in the table, including single parameter and multiple parameters. The performance increased when including multiple parameters compared to a single input. Basic single signal representations performed poorly: log-compressed (B-mode) data yielded the weakest performance ($R_s = 0.18$, AUC = 0.62), while envelope and raw RF data offered only modest improvement ($R_s = 0.46$; AUC = 0.67 and 0.73, respectively). In contrast, the physics driven QUS parameters individually achieved improved performance. H-scan ($R_s = 0.90$, AUC = 0.96) and attenuation ($R_s = 0.89$, AUC = 0.96) were the strongest single parameters, followed by Nakagami (Rs = 0.83, AUC = 0.92). Combining physics driven parameters further improved performance. Adding raw RF data to these QUS parameters provided only limited additional benefit (e.g., RF + H-scan: Rs = 0.89, AUC = 0.95; RF + H-scan + Attenuation: $R_s = 0.90$, AUC = 0.97; RF + H-scan + Attenuation + Nakagami: $R_s = 0.90$, AUC = 0.97); we included RF data, since RF data performed the best among the echogenicity-based representations (RF, envelope, and log-compressed data). The best-performing configuration combined the three QUS parameters (H-scan, attenuation, and Nakagami), achieving the high overall performance ($R_s = 0.92$, AUC = 0.98). This combination outperformed each parameter individually as well as every RF-inclusive configuration.

**Table 1.** Performance of individual and combined QUS input channels for predicting liver fat fraction using our deep learning framework, evaluated against MRI-PDFF as the reference standard. For classification, MRI-PDFF cutoffs of 6.4%, 17.4%, and 22.1% were used to define steatosis grade boundaries (S0/S1, S1/S2, S2/S3, respectively)

| **Input Combinations** | **$R_s$** | **AUC** |
|---|---|---|
| Log-Compressed | 0.18 | 0.62 |
| Envelope | 0.46 | 0.67 |
| Radio Frequency (RF) Data | 0.46 | 0.73 |
| H-Scan | 0.90 | 0.96 |
| Attenuation | 0.89 | 0.96 |
| Nakagami | 0.83 | 0.92 |
| RF + H-scan | 0.89 | 0.95 |
| RF + H-scan + Attenuation | 0.90 | 0.97 |
| RF + H-scan + Attenuation + Nakagami | 0.90 | 0.97 |
| **H-scan + Attenuation + Nakagami** | **0.92** | **0.98** |

**Figure 1** shows the correlation between the fat fraction predicted by the best-performing model (H-scan + attenuation + Nakagami input) and the reference MRI-PDFF, with each data point representing a single patient-level prediction averaged across multiple scans. The predicted values closely tracked the reference values along the line fit, particularly within the fat fraction range containing the majority of cases (MRI-PDFF < 30%), where the fitted line nearly coincided with the identity relationship, indicating accurate and unbiased estimation across the normal-to-severe steatosis spectrum. Only in the extreme high-fat-range (MRI-PDFF > 30%) did the model tend to underestimate fat fraction. The tight clustering of points around the line is consistent with the high Spearman correlation ($R_s$ = 0.92).

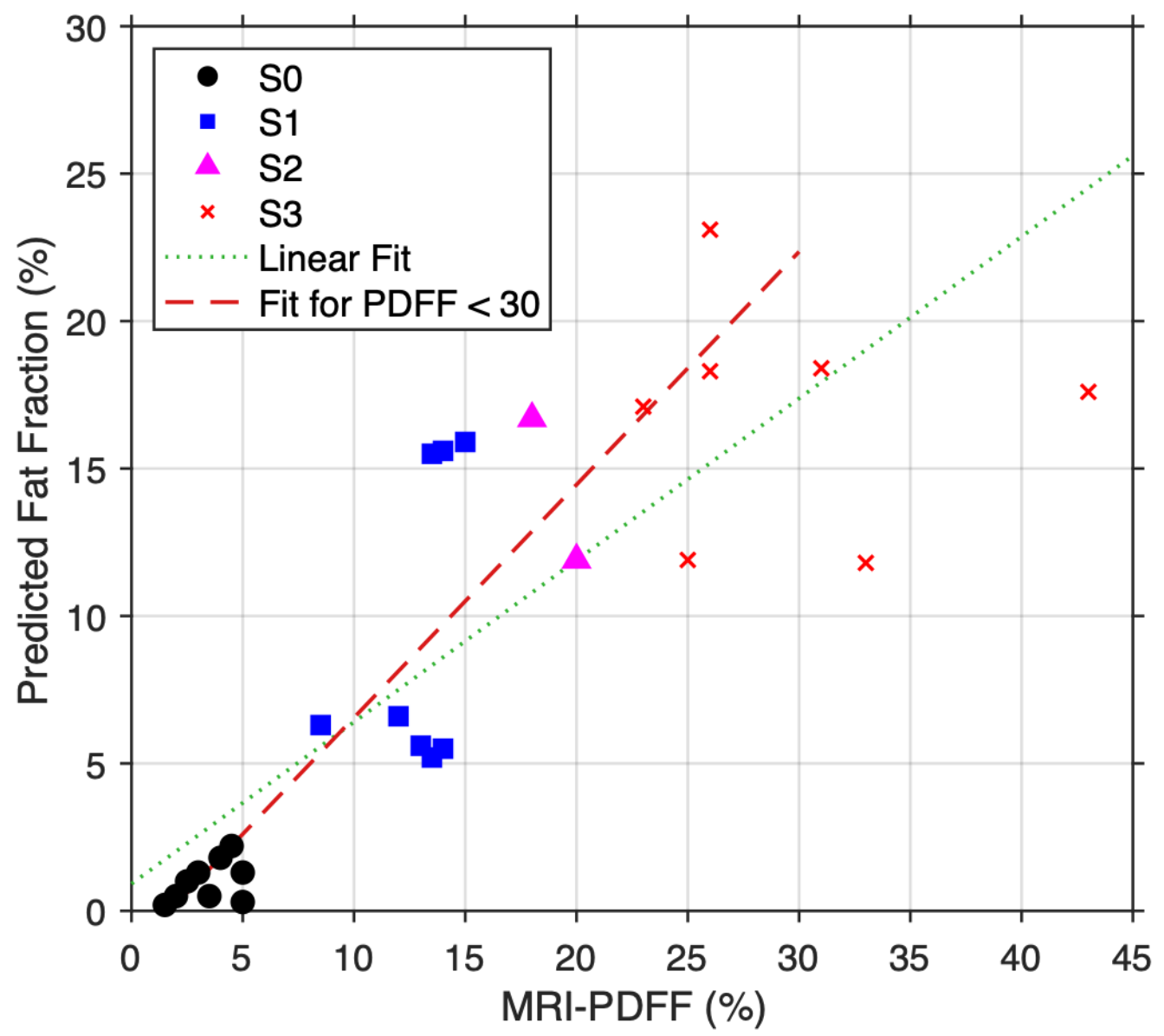

**Figure 1.** Correlation between predicted fat fraction (H-scan + attenuation + Nakagami input) and reference MRI-PDFF. Each data point represents a single patient prediction averaged across multiple scans.

**Figure 2** shows representative multiparametric imaging results across patients spanning a range of increasing MRI-PDFF values (reference fat fractions of approximately 1.25% to 25.9%). For each case, the corresponding B-mode image, the predicted pixel-wise fat fraction map, and the prediction overlay are shown. As the reference MRI-PDFF increased, the predicted fat fraction maps exhibited a progressive color shift from blue (low fat) toward yellow and red (high fat), demonstrating clear concordance with gradual fat accumulation. In the lowest-fat case (reference, predicted = 1.25%, 0.12%), the predicted map was uniformly blue, whereas in the highest-fat case (25.875%, 24.41%), the map displayed widespread elevated fat fraction across the liver region. Importantly, the predicted maps revealed spatially heterogeneous fat distribution within individual livers rather than a single uniform value, with regions of locally elevated fat accumulation visible in intermediate cases. The close numerical agreement between reference and predicted values across the displayed cases, together with the strong spatial correspondence seen in the prediction overlays, demonstrates that the model produces both accurate patient-level estimates and spatially coherent, pixel-wise fat fraction maps.

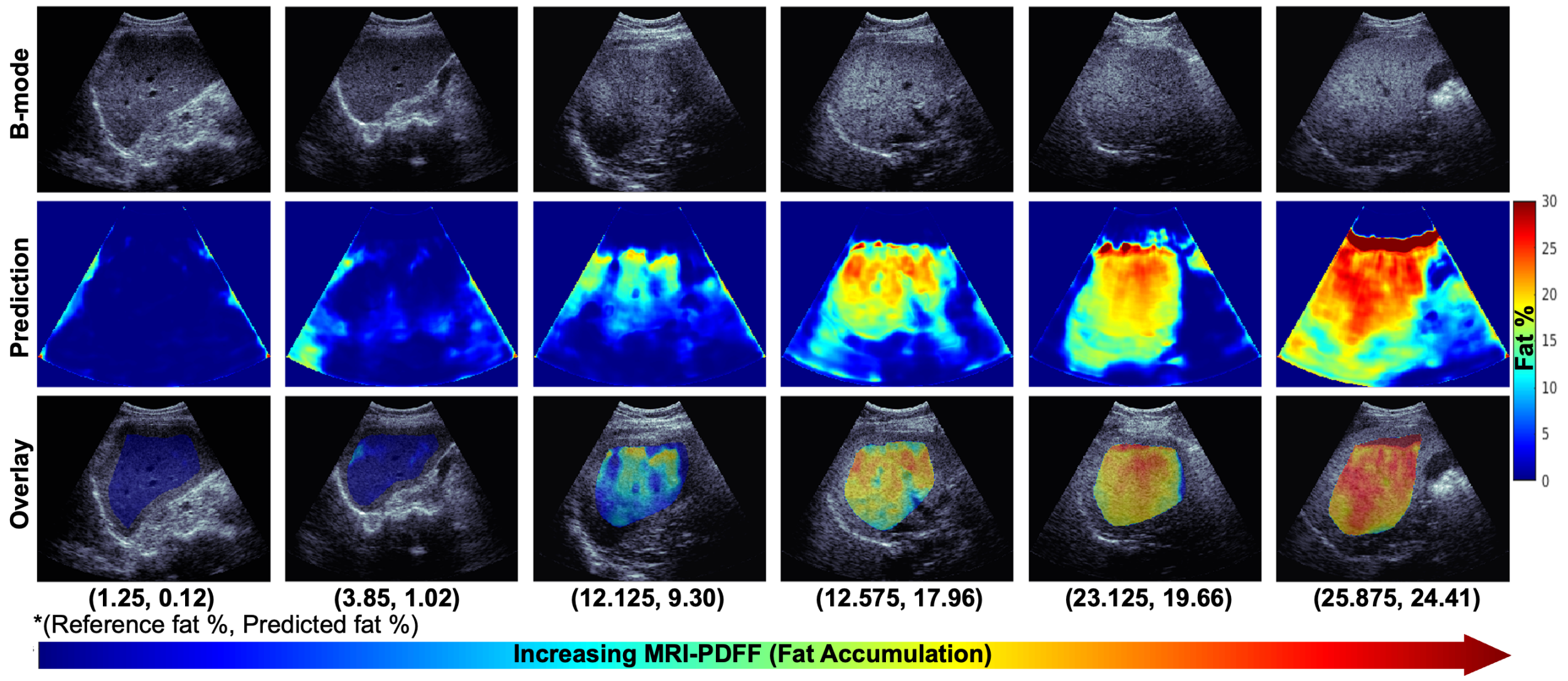

**Figure 2.** Multiparametric imaging visualizing estimated fat fraction (%) on a pixel-wise basis. Each column corresponds to a representative case with increasing hepatic fat accumulation (left to right). Top row: B-mode ultrasound images. Middle row: predicted fat fraction maps. Bottom row: predicted fat fraction within the liver overlaid on the corresponding B-mode image. The colorbar represents fat %, ranging from 0 to 30%. Values below each column denote (reference fat fraction, predicted fat fraction)

## 4. Discussion

This study demonstrates that a deep learning-based multiparametric QUS framework can accurately estimate hepatic fat fraction from ultrasound RF-derived parameter maps, achieving strong agreement with the MRI-PDFF reference standard ($R_s = 0.92$, AUC = 0.98) while producing spatially resolved, pixel-wise fat fraction maps.

The network performance depended strongly on the input configuration, with the physics-driven QUS configurations outperforming the basic echogenicity-based configurations by a large performance gap. Log-compressed, envelope, and raw RF data each yielded poor-to-modest predictive performance ($R_s \leq 0.46$), whereas H-scan, attenuation, and Nakagami parameters individually achieved strong performance ($R_s = 0.83$–$0.90$). This large performance gap highlights the importance of physics-driven feature extracting, including spectral analysis, attenuation estimation, and speckle statistics, for feeding the network, rather than relying on the network's own feature extraction, particularly since all configurations were derived from the same underlying RF data. Although deep networks are theoretically capable of learning relevant features directly from raw data, the limited size of clinical datasets (here, 30 patients) may constrain the ability. Providing the network with physics-driven QUS parameters that already encode meaningful tissue scattering and acoustic attenuation properties therefore serves as an informative prior and substantially improves performance.

The fact that combining H-scan, attenuation, and Nakagami parameters ($R_s = 0.92$, AUC = 0.98) outperformed each parameter individually indicates that these parameters carry complementary, non-redundant information about the underlying tissue signatures, which the network integrates in a data-driven manner. In contrast, the addition of raw RF data did not further improve performance and was unnecessary in the best configuration, suggesting that once the informative parameters are provided, the raw data contributes little beyond redundancy.

As shown in **Figure 1**, the predicted fat fraction tracked the reference MRI-PDFF closely across the normal-to-severe steatosis spectrum, with the fitted line nearly coinciding with the identity relationship for the majority of cases (MRI-PDFF < 30%). Underestimation occurred only in the extreme high-fat range (MRI-PDFF > 30%). These deviations likely reflect the limited number of high-steatosis cases available for training.

Although our ground truth is a single value MRI-PDFF for each whole liver, the network successfully generated pixel-wise fat fraction maps that capture heterogeneous fat accumulation within the liver. To enable this localized prediction, we extracted random size patches more than feeding the network using the whole liver images, which were used to train the network. Moreover, our hybrid loss function included RMSE to compare the averaged prediction within each patch with MRI-PDFF, in addition to the pixel-wise comparison by the cross entropy. As shown in **Figure 2**, as the reference MRI-PDFF increased, the predicted maps exhibited a progressive color shift from blue to yellow and red, demonstrating clear concordance with gradual fat accumulation. Importantly, rather than reporting a single whole-liver value, as is the commercial ultrasound derived fat fraction (UDFF) by Siemens, our framework reveals the spatial distribution of fat, including regions of locally elevated accumulation in intermediate cases. This capability may be clinically valuable for characterizing non-uniform liver involvement and for localizing regions of early or focal disease.

Our approach achieved performance (AUC = 0.98) comparable to our previous PCA-based linear multiparametric QUS method [18] and a single-channel deep learning-based classification [21], both of which reported AUC values greater than 0.98. All three studies were evaluated on the same human dataset, enabling a direct comparison. The principal advantage of the proposed framework, however, lies in its pixel-wise, continuous fat fraction prediction. Baek et al. [18] produced pixel-wise imaging of local fat fraction but provided very limited differentiation within a liver, resulting in nearly homogeneous predictions. In contrast, our framework generates continuous fat fraction maps in which color varies smoothly with local fat content, thereby representing the spatial distribution and heterogeneity of fat accumulation within the liver rather than a near-uniform classification. Moreover, Sanabria et al. [21] predicted discrete steatosis stages (0 to 3), whereas our network directly predicts the quantitative fat fraction, providing finer-grained and more interpretable estimates despite a comparable AUC of 0.98. While their work employed a CNN classifier, we developed our steatosis estimation network based on the more spatially aware UNet architecture [23] to preserve fine spatial detail through skip connections, achieving localized, high-resolution fat fraction maps.

To further improve imaging quality, more advanced, higher performance architectures such as transformers [25, 26] could be adopted. In addition, to rigorously validate the pixel-wise imaging accuracy of our predictions, it would be valuable to acquire additional human subjects imaged in views that capture more than one liver lobe with differing amounts of accumulated fat, enabling direct assessment of intra-liver spatial accuracy. We would further validate our framework on more complex diseases characterized by heterogeneous progression within the liver, such as metabolic dysfunction-associated steatohepatitis (MASH), in which inflammation, steatosis, and fibrosis progress simultaneously and heterogeneously throughout the liver, conditions for which the spatially resolved, continuous prediction offered by our framework is expected to be particularly advantageous.

The relatively small cohort (n = 30) limits statistical power and may restrict generalizability, particularly for severe steatosis, which was underrepresented, as reflected by the outliers at high MRI-PDFF values in **Figure 1**. Larger with diverse patient populations will be needed to validate and generalize these findings. In addition, the current framework focuses on steatosis quantification; extending it to MASH, where the ability to capture spatial heterogeneity is expected to be most advantageous, remains an important direction for future work.

## 5. Conclusion

We proposed a deep learning-based multiparametric QUS framework that integrates multiple physics-driven QUS parameters to estimate pixel-wise hepatic fat fraction, using a U-Net architecture trained with MRI-PDFF as the reference standard. The input configuration of H-scan, attenuation, and Nakagami parameters achieved strong agreement with MRI-PDFF ($R_s = 0.92$, AUC = 0.98), outperforming both single-parameter inputs and RF-inclusive configurations. The findings demonstrate that physics-driven feature extraction and the integration of complementary, non-redundant QUS parameters are key to accurate fat fraction estimation. Overall, we successfully implemented the multiparametric QUS analysis concept within a deep learning framework. Crucially, our framework moves beyond conventional single-value reporting by generating continuous, spatially resolved fat fraction maps that capture heterogeneous fat distribution within the liver. We anticipate that our approach will offer a clinically valuable tool for characterizing non-uniform liver disease development and localizing early or focal disease, and may prove particularly advantageous for assessing more complex, heterogeneously progressing liver diseases such as metabolic dysfunction-associated steatohepatitis (MASH).

*Acknowledgments*— This work was supported Radiological Sciences Laboratory (RSL) program and the School of Medicine Dean's Postdoctoral Fellowship, Stanford University. The authors also acknowledge Canon Medical Systems for providing funding support, equipment, and resources for this project.

## References

[1] Z. M. Younossi, P. Golabi, J. M. Paik *et al.*, "The global epidemiology of nonalcoholic fatty liver disease (NAFLD) and nonalcoholic steatohepatitis (NASH): a systematic review," *Hepatology,* vol. 77, no. 4, pp. 1335-1347, 2023.

[2] C. Estes, H. Razavi, R. Loomba *et al.*, "Modeling the epidemic of nonalcoholic fatty liver disease demonstrates an exponential increase in burden of disease," *Hepatology,* vol. 67, no. 1, pp. 123-133, 2018.

[3] Z. M. Younossi, M. Stepanova, Y. Younossi *et al.*, "Epidemiology of chronic liver diseases in the USA in the past three decades," *Gut,* vol. 69, no. 3, pp. 564-568, 2020.

[4] M. E. Rinella, B. A. Neuschwander-Tetri, M. S. Siddiqui *et al.*, "AASLD Practice Guidance on the clinical assessment and management of nonalcoholic fatty liver disease," *Hepatology (Baltimore, Md.),* vol. 77, no. 5, pp. 1797, 2023.

[5] W. Eskridge, D. R. Cryer, J. M. Schattenberg *et al.*, "Metabolic dysfunction-associated steatotic liver disease and metabolic dysfunction-associated steatohepatitis: the patient and physician perspective," *Journal of Clinical Medicine,* vol. 12, no. 19, pp. 6216, 2023.

[6] M. E. Rinella, J. V. Lazarus, V. Ratziu *et al.*, "A multisociety Delphi consensus statement on new fatty liver disease nomenclature," *Hepatology,* vol. 78, no. 6, pp. 1966-1986, 2023.

[7] E. C. Lee, V. V. Anand, A. C. Razavi *et al.*, "The global epidemic of metabolic fatty liver disease," *Current Cardiology Reports,* vol. 26, no. 4, pp. 199-210, 2024.

[8] P. Angulo, D. E. Kleiner, S. Dam-Larsen *et al.*, "Liver fibrosis, but no other histologic features, is associated with long-term outcomes of patients with nonalcoholic fatty liver disease," *Gastroenterology,* vol. 149, no. 2, pp. 389-397. e10, 2015.

[9] A. J. Sanyal, M. L. Van Natta, J. Clark *et al.*, "Prospective study of outcomes in adults with nonalcoholic fatty liver disease," *New England Journal of Medicine,* vol. 385, no. 17, pp. 1559-1569, 2021.

[10] A. M. Pirmoazen, A. Khurana, A. El Kaffas *et al.*, "Quantitative ultrasound approaches for diagnosis and monitoring hepatic steatosis in nonalcoholic fatty liver disease," *Theranostics,* vol. 10, no. 9, pp. 4277-4289, 2020.

[11] J. Neuberger, J. Patel, H. Caldwell *et al.*, "Guidelines on the use of liver biopsy in clinical practice from the British Society of Gastroenterology, the Royal College of Radiologists and the Royal College of Pathology," *Gut,* vol. 69, no. 8, pp. 1382-1403, 2020.

[12] N. Pineda, P. Sharma, Q. Xu *et al.*, "Measurement of hepatic lipid: high-speed T2-corrected multiecho acquisition at 1H MR spectroscopy—a rapid and accurate technique," *Radiology,* vol. 252, no. 2, pp. 568-576, 2009.

[13] T. Yokoo, M. Shiehmorteza, G. Hamilton *et al.*, "Estimation of hepatic proton-density fat fraction by using MR imaging at 3.0 T," *Radiology,* vol. 258, no. 3, pp. 749-759, 2011.

[14] S. B. Reeder, I. Cruite, G. Hamilton *et al.*, "Quantitative assessment of liver fat with magnetic resonance imaging and spectroscopy," *Journal of magnetic resonance imaging,* vol. 34, no. 4, pp. 729-749, 2011.

[15] A. M. Pirmoazen, A. Khurana, A. M. Loening *et al.*, "Diagnostic performance of 9 quantitative ultrasound parameters for detection and classification of hepatic steatosis in nonalcoholic fatty liver disease," *Investigative Radiology,* vol. 57, no. 1, pp. 23-32, 2022.

[16] K. Parker, "Scattering and reflection identification in H-scan images," *Physics in Medicine & Biology,* vol. 61, no. 12, pp. L20, 2016.

[17] K. J. Parker, and J. Baek, "Fine-tuning the H-scan for discriminating changes in tissue scatterers," *Biomedical physics & engineering express,* vol. 6, no. 4, pp. 045012, 2020.

[18] J. Baek, A. El Kaffas, A. Kamaya *et al.*, "Multiparametric quantification and visualization of liver fat using ultrasound," *WFUMB Ultrasound Open,* vol. 2, no. 1, pp. 100045, 2024.

[19] J. Baek, L. Basavarajappa, R. Margolis *et al.*, "Multiparametric ultrasound imaging for early-stage steatosis: Comparison with magnetic resonance imaging-based proton density fat fraction," *Medical physics,* vol. 51, no. 2, pp. 1313-1325, 2024.

[20] T.-Y. Huang, Z.-Y. Li, J. Tian *et al.*, "Metabolic Dysfunction–Associated Steatotic Liver Disease at Quantitative US: International Prospective Study," *Radiology,* vol. 316, no. 1, pp. e242564, 2025.

[21] S. J. Sanabria, A. M. Pirmoazen, J. Dahl *et al.*, "Comparative Study of Raw Ultrasound Data Representations in Deep Learning to Classify Hepatic Steatosis," *Ultrasound in Medicine & Biology,* vol. 48, no. 10, pp. 2060-2078, 2022.

[22] P. M. Shankar, "A general statistical model for ultrasonic backscattering from tissues," *IEEE transactions on ultrasonics, ferroelectrics, and frequency control,* vol. 47, no. 3, pp. 727-736, 2000.

[23] O. Ronneberger, P. Fischer, and T. Brox, "U-net: Convolutional networks for biomedical image segmentation." pp. 234-241.

[24] M. Abadi, P. Barham, J. Chen *et al.*, "{TensorFlow}: a system for {Large-Scale} machine learning." pp. 265-283.

[25] A. Vaswani, N. Shazeer, N. Parmar *et al.*, "Attention is all you need," *Advances in neural information processing systems,* vol. 30, 2017.

[26] A. Dosovitskiy, "An image is worth 16x16 words: Transformers for image recognition at scale," *arXiv preprint arXiv:2010.11929*, 2020.